\documentclass[epj]{svjour}
\usepackage{amsmath}
\usepackage{graphicx}
\begin{document}

\title{Percolation of the aligned dimers on a square lattice }

\author{V.A.~Cherkasova\inst{1} \and  Yu.Yu.~Tarasevich\inst{1} \and N.I.~Lebovka\inst{2} \and N.V.~Vygornitskii\inst{2}}

\institute{Astrakhan State University, 20a Tatishchev Str, Astrakhan,  414056,
Russia \and Institute of Biocolloidal Chemistry named after F.D. Ovcharenko,
NAS of Ukraine, 42, blvr. Vernadskogo, Kyiv, 03142, Ukraine}
\date{Received: date / Revised version: date}
%
\abstract{Percolation and jamming phenomena are investigated for anisotropic
sequential deposition of dimers (particles occupying two adjacent adsorption
sites) on a square lattice. The influence of dimer alignment on the electrical
conductivity was examined. The percolation threshold for deposition of dimers
was lower than for deposition of monomers. Nevertheless, the problem belongs to
the universality class of random percolation. The lowest percolation threshold
($p_c=0.562$) was observed for isotropic orientation of dimers. It was higher
($p_c=0.586$) in the case of dimers aligned strictly along one direction. The
state of dimer orientation influenced the concentration dependence of
electrical conductivity. The proposed model seems to be useful for description
of the percolating properties of anisotropic conductors.
} 
\maketitle

\section{Introduction}\label{sec:introduction}

Physical properties of (partially) disordered systems are described in
percolation
approach~\cite{Grimmet,Kesten,Feder,Ziman,Shklovskii,Stauffer,Sahimi}. Effect
of concentration on physical properties is well understood. Nevertheless,
alignment effect plays essential role for high aspect ratio objects, such as
nanotubes and nano\-wires~\cite{Du}.

Percolating properties and behaviors of systems, composed of aniso\-tropic
nanoparticles, are extensively investigated during last
years~\cite{Carroll,Sreekumar,Zhou,dovzhenko,Kondrat,Lebovka}. The problem of
aniso\-tropic percolation has been a subject of many investigations. It is of
interest and of value to inquire how an orientation of nanoparticles influences
the main physical properties of the systems. Effect of the nanotube alignment on
conductivity is of specific interest. The anisotropy can be induced by various
factors. For instance, alignment of the nanotubes may be induced by flow~\cite{Du} or
electric field~\cite{Liu,Park}.

The electrical properties of a ceramic composition were considered
in~\cite{dovzhenko}. A profound influence of the conducting phase structure on
the electrical conductivity of the material was shown. The simulations indicated
considerable dependence of the percolation cluster properties on
anisotropy of its components.

The irreversible adsorption (deposition) of particles on solid
surfaces is a subject of considerable practical importance. A well
known example of an irreversible monolayer deposition process is
the random sequential adsorption (RSA). RSA has attracted
significant interest due to its importance in many physical,
chemical, and biological processes. RSA is a natural model for
irreversible and sequential deposition of macromolecules at
solid--liquid interfaces. Some examples of the wide range of
applicability of this model include adhesion of colloidal
particles, as well as adsorption of proteins to solid surfaces,
with relaxation times much longer than the deposit formation time.
This process is well described in the literature and has been
investigated extensively in the last decades. The topic has been
well covered 
in~\cite{Evans,Adamczyk}.

Investigation of the irreversible adsorption of polyatomic species
($k$-mers) has received considerable attention in the last years.
The results of the study of random sequential adsorption and
percolation of polyatomic species on different substrates were
presented in~
\cite{Kondrat,Holloway,Evans1989,Leroyer,Vandewalle,Cortes,Cornette1,Cornette2,Cornette3,Quintana,Dolz}.
Numerical studies of random sequential adsorption (the RSA model)
of rectangular particles on a flat substrate are performed and
dependencies of the saturation concentration and the percolation
threshold on the model parameters are determined
in~\cite{Lebovka}.

The main goal of the present study is to investigate the influence
of alignment on the percolation and jamming thresholds, as well as
conductivity. In this paper, we provide accurate numerical data
for the percolation and jamming thresholds of the (partially)
ordered (aligned) dimers on a square lattice.
We perform calculation of electrical conductivity
as a function of dimers alignment.
The proposed model can be useful for description of the percolation behaviour of
anisotropic conductor networks.

The paper is organized as follows. In Section~\ref{sec:model} the
basis of the model of deposition of dimers on a square lattice is
presented. of The results, obtained using the finite size scaling
theory, are also analysed and discussed in this Section. The
results, related to conductivity of the systems concerned, are
presented in Section~\ref{sec:conductivity} pres. Finally, we
discuss dependence of the percolation threshold on the model
parameters of interest in Section~\ref{sec:conclusion}.

\section{Model and Simulations}\label{sec:model}
Different kinds of boundary conditions were used in percolation
simulations: open, periodic or toroidal, and cylindrical, i.e.
open along one direction and periodic along another one. One uses
usually the term "crossing" only to systems with open boundaries,
and the terms "spanning" and "wrapping" for cylindrical boundary
conditions~\cite{Pruessner}. In our study, the periodic boundary
conditions are applied by gluing along the vertical and horizontal
borders. We define a wrapping cluster as a cluster that winds
around the system along the given direction, i.e. it provides a
path with the length of $2\pi$~\cite{Pruessner}.

Mersenne Twister random number generator~\cite{Matsumoto} was utilized for
filling in the lattice with dimer at given concentration and orientation. It has
a period of $2^{19937} - 1$.

The anisotropy of dimer orientation is described by the orientation order parameter $s$:
$$
s = \frac{N_{|}-N_{-}}{N_{|}+N_{-}}, $$ where $N_{|}$ and  $N_{-}$ are the
numbers of dimers, oriented in vertical and horizontal direction,
respectively~\cite{Matoz-Fernandez}.

Let us consider a periodic square lattice of linear size $L$, on which dimers are deposited at random, but with given orientation.
Two of nearest neighbour
sites along given direction are randomly selected; if both sites are vacant, the
dimer adsorbs on such sites. Otherwise, the attempt is rejected. In
any case, the procedure is iterated until $N = N_| +N_-$ dimers are adsorbed
and the desired concentration is reached.

We perform our calculations of percolation thresholds using the
Hoshen--Kopelman algorithm~\cite{Hoshen}.

\begin{figure}[!htbp]
\centering\includegraphics*[width=0.5\linewidth]{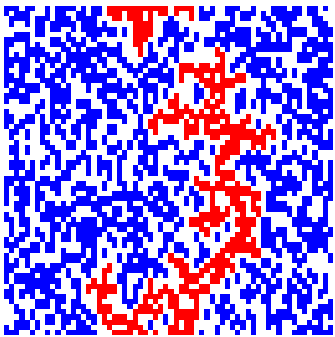}\hfill \caption {
Color online. Percolation on a square lattice of linear size $L=64$ at $s = 1$
(wrapping cluster is indicated in red)}\label{fig:L05}
\end{figure}

Fig.~\ref{fig:L05} demonstrates percolation in an anisotropic system of dimers
(the dimers are oriented strictly along the vertical direction).

The final state, generated by irreversible adsorption, is a disordered state
(known as jamming state), in which no more objects can be deposited due to the
absence of any free space of appropriate size and shape~\cite{Evans}. If
different orientations of the deposited objects are not equally probable, the
definition of the jamming state is to be refined. Let us assume $N_{|} > N_{-}$.
We define jamming for the fixed parameter $s$ as a situation when there is no
possibility of depositing any additional vertically oriented object.
Nevertheless, there may be places for accepting horizontally oriented objects.

\begin{figure}[!htbp]
\includegraphics*[width=0.45\linewidth]{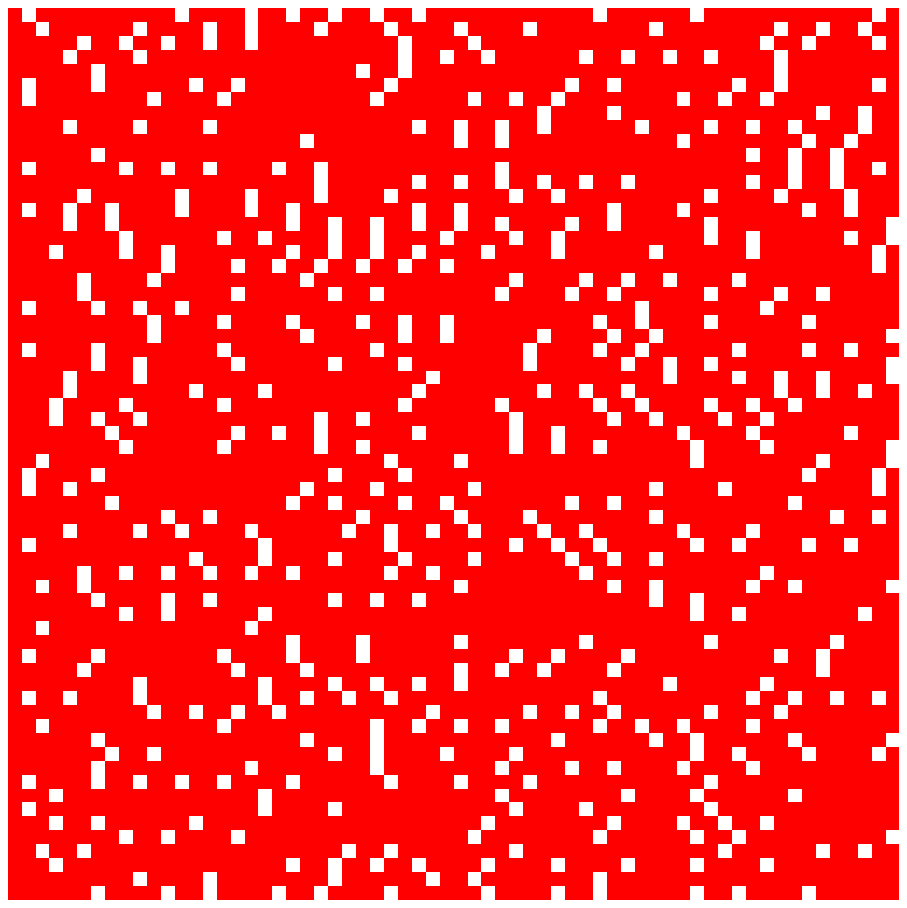}\hfill\includegraphics*[width=0.45\linewidth]{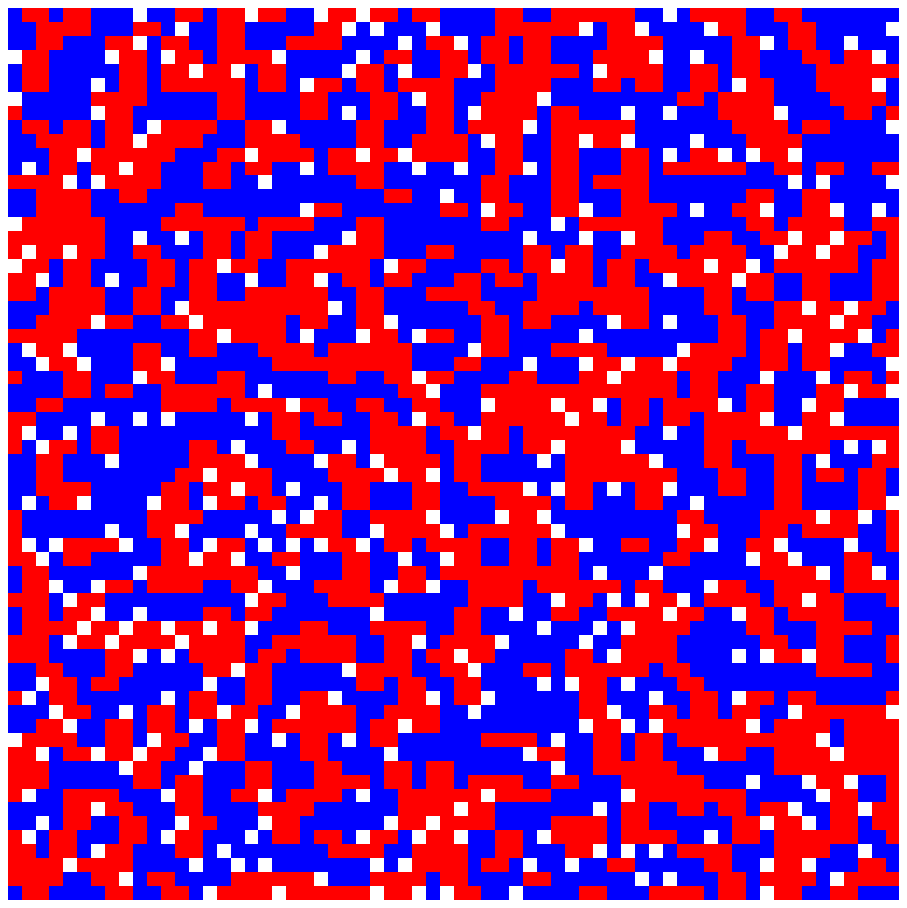}
\caption{Color online. Jamming at a lattice of linear size $64$ at $s = -1$
(left) and $s = 0$ (right) (horizontally oriented dimers are shown in red,
whereas the vertically oriented objects are shown in blue )}\label{fig:L0j}
\end{figure}

Fig.~\ref{fig:L0j} shows jamming for different $s$ at a lattice of linear size
$L = 64$.

\begin{figure}[!htbp]
\centering\includegraphics*[width=0.7\linewidth]{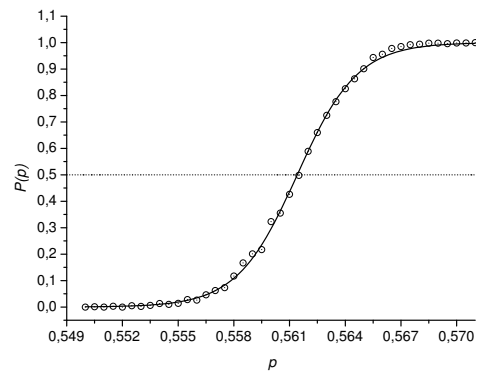} \caption
{Percolation probability $P$ vs. occupation probability $p$ for a lattice of
linear size $L=1024$ at $s = 0.4$, $p_c = 0.56141 \pm 0.00003$}\label{fig:oc03}
\end{figure}

The percolation probability $P$ vs. the site occupation probability $p$ was
obtained for the lattices of linear sizes $L = 64, 128, 256, 512$, $1024$ and
the number of runs $1000$. The percolation probability $P$ as a function of occupation $p$ for a particular value of $s$ is shown in
Fig.~\ref{fig:oc03}. The percolation threshold $p_c$ for a given lattice size $L$ can be estimated from the condition $P(p)=0.5$.

\begin{figure}[!htbp]
\centering\includegraphics*[width=0.7\linewidth]{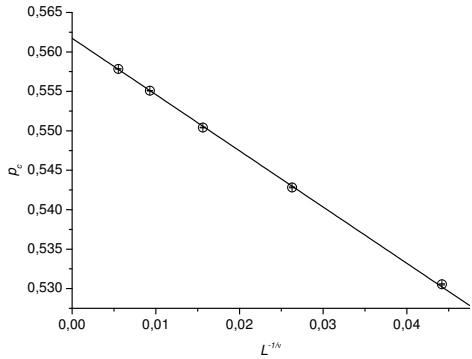}\caption {Example of
percolation threshold determination in the thermodynamical limit of ($L \to
\infty$) using the scaling relation~\eqref{eq:scal} ($s = 0$, $p_c = 0.56173
\pm 0.00003$)}\label{fig:sc05}
\end{figure}

The finite-size scaling analysis was carried out for getting the percolation
threshold at $L \to \infty$. The percolation threshold $p_c(L)$ was calculated
for five different values of the linear lattice size. The percolation threshold
$p_c(\infty)$ of an infinite lattice can be found by fitting the results for
lattices of different sizes to the scaling relation (Fig.~\ref{fig:sc05}):
\begin{equation}
   \left| p_c(L) - p_c(\infty) \right|  \propto  L^{-1/\nu},
   \label{eq:scal}
\end{equation}
where the critical exponent $\nu$ has the value $4/3$ in three
dimensions~\cite{Stauffer}.

In particular, our simulations gave the percolation threshold $p_c = 0.5863$ for
strictly oriented dimers ($s = 1$), and  $p_c = 0.5617$ for dimers oriented in
two directions with equal probability ($s = 0$). The last result agrees with 
the known results~\cite{Vandewalle,Dolz} (see Table.~\ref{tab:pc_comp}).

\begin{table}[!htbp]
  \centering
  \caption{Comparison of published and our results for dimers
oriented in two directions with equal probability}\label{tab:pc_comp}
  \begin{tabular}{|p{3.5cm}|c|c|c|}
    \hline
        & $p_c$ & $p_{jam}$ & $p_c/p_{jam}$ \\ \hline
    Our result & $0.5617$ & $0.90668$ & $0.6195$ \\ \hline
     Open boundary conditions, $L_\text{max} =
2000$~\cite{Vandewalle}& $0.562$ & $0.906$ & $0,62 \pm 0.01$
\\ \hline Periodic boundary conditions, $L_\text{max} = 112$, $5\times10^4$ runs~\cite{Dolz}
 & $0.562$ & $0.907$ & $0.6196$ \\
    \hline
  \end{tabular}
\end{table}

Moreover, several quantities of interest ~\cite{Stauffer} were calculated.

Average cluster size:
\begin{equation}
S~= \sum_{i}n_i i^2/\sum_{i}n_i i,
  \label{eq:av}
\end{equation}
where $n_i$ is the average number of $i$ clusters per lattice site.

The strength of the infinite network $P_{\infty}$, i.e. the probability of an
arbitrary site belonging to the infinite network:
\begin{equation}
P_{\infty}(p)~= \frac{N_{\infty}}{N},
  \label{eq:pinf}
\end{equation}
where $N$ is the total number of sites.

The average cluster size $S$ and strength of the infinite network $P_\infty$
yield the scaling laws:
\begin{equation}
   S(p) \propto \left| p - p_c \right|^{-\gamma},
   \label{eq:ssc}
\end{equation}
and
\begin{equation}
   P_\infty(p) \propto \left( p - p_c \right)^{\beta},
   \label{eq:be}
\end{equation}
where $\gamma$ and  $\beta$ the universal critical exponents. In $d =
2$~\cite{Bunde}:
\begin{equation}
 \gamma = 2.3889,
   \label{eq:gam}
\end{equation}
and
\begin{equation}
 \beta = 0.1389.
   \label{eq:bet}
\end{equation}

In our computations, the average cluster size~\eqref{eq:av} and the strength of
the infinite network~\eqref{eq:pinf} demonstrate typical behavior near the
percolation transition. The critical exponents $\gamma = 2.02 \pm 0.01$, at
$p < p_c$, $\gamma = 2.37 \pm 0.05$, when $p
> p_c$, and $\beta = 0.179 \pm 0.002$ for $s = 0$, extracted from the power
laws~\eqref{eq:ssc} and~\eqref{eq:be}, are close to the known
values~\eqref{eq:gam} and~\eqref{eq:bet}.

The fractal dimension of the incipient cluster~\cite{Stauffer}, calculated as
$$
 d_f = d - \frac{\beta}{\nu}
$$,
is close to the known values $d_f = 1.896$. For instance, $d_f = 1.8896$ for
$s = 0$ .

\begin{figure}[!htbp]
\centering
\includegraphics*[width=0.9\linewidth]{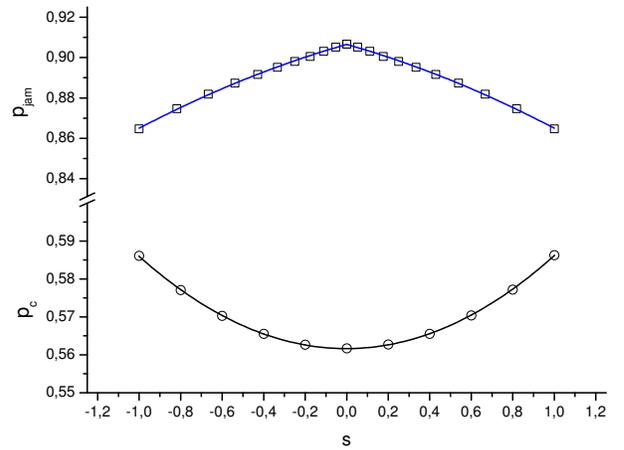}\caption {
Percolation threshold $p_c$ vs. orientation order parameter $s$, $p_c(s) = 0.02442s^2 + 0.56165$, and
jamming threshold vs. $s$, $p_{jam}(s) = -0.0124s^2 - 0.0291|s| + 0.9065$
}\label{fig:graphj}
\end{figure}

We found that parabolic law is a reasonable fit for $p_c$ vs. $s$
(Fig.~\ref{fig:graphj}).

The jamming threshold for $L \to \infty$ was found from scaling
relation~\eqref{eq:scal}, where $\nu = 1.0 \pm 0.1$ is the critical
exponent~\cite{Vandewalle}. In particular, for strictly oriented dimers ($s =
\pm 1$) (in fact, this case is quite equivalent to jamming in one dimension)
$p_{jam} = 0.8646$, while exact value for jamming in one dimension is $p_{jam}
= 1 - e^{-2} \approx 0.86466$~\cite{Evans}. Fig.~\ref{fig:graphj} demonstrates
$p_\text{jam}(s)$.

\section{Electrical conductivity}\label{sec:conductivity}

The highly efficient algorithm, proposed by Frank and Lobb~\cite{FL88} was
utilized for finding conductivity of a square lattice, filled with the dimers.

\begin{figure}[!htbp]
\centering
\includegraphics*[width=0.7\linewidth]{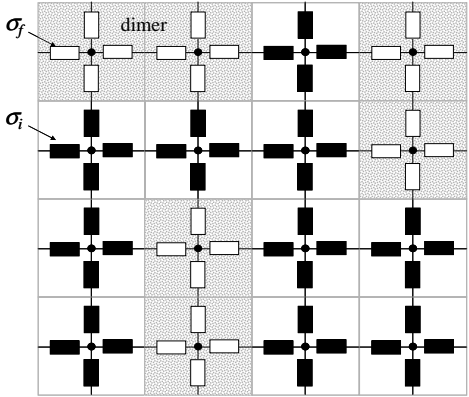}
\caption{Equivalent schema of a square lattice filled with dimers, where
$\sigma_f$ and $\sigma_i$ are conductivity of the occupied side
($\sigma_f=10^6$) and empty one ($\sigma_i=1$), respectively}\label{fig:Leb01}
\end{figure}

The Frank and Lobb algorithm utilises the repeated application of a sequence
of series, parallel and star-triangle (Y--$\triangle$) transformations to the
bonds of the lattice. The final result of this sequence of transformations is
reduction of any finite portion of the lattice to a single bond that has the
same conductance as the entire lattice. We used four equivalent resistors
(conductors) with $\sigma_f=10^6$ and $\sigma_i=1$ for occupied and empty sites
\cite{Yuge78}, respectively, instead of each cell (see Fig.~\ref{fig:Leb01}).

\begin{figure}[!htbp]
\centering
\includegraphics*[width=0.8\linewidth]{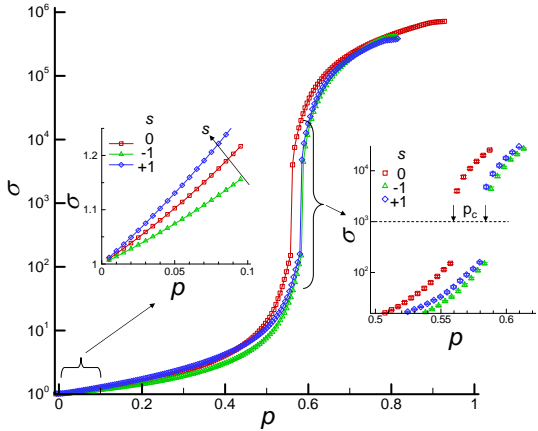}\caption {
Conductivity $\sigma$ vs. occupation probability $p$ for isotropic ($s =
0$)and strictly oriented dimers along the fixed direction ($s = \pm 1$). The
lattice linear size is $L = 256$}\label{fig:Leb02}
\end{figure}

Examples of conductivity $\sigma$ as a function of occupation probability   $p$
at different values of parameter $s$ are shown in Fig.~\ref{fig:Leb02}.
Behavior of the conductivity $\sigma(p)$ corresponds with direct estimations of the
percolation threshold  $p_c$. There is a sharp transition of conductivity from
$\sigma_-$ to $\sigma_+$ near the percolation threshold (see right inset in
Fig.~\ref{fig:Leb02}). Moreover, $\sigma$ increases on passing from isotropic
($s = 0$) to strictly ordered dimers ($s = \pm 1$).

\begin{figure}[!htbp]
\centering
\includegraphics*[width=0.7\linewidth]{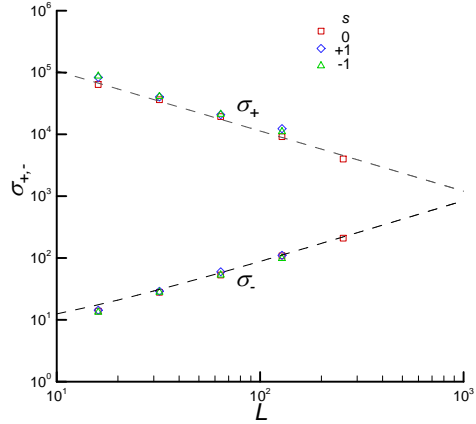}\caption {
Application of scaling to calculation of the ratio of conductivity exponent
to the exponent of the correlation length for isotropic ($s=0$) and strictly
oriented ($s= \pm 1$) dimers. Dashed line corresponds to the slope 0.973, which
is typical for percolation in two dimensions~\cite{LF84}}\label{fig:Leb03}
\end{figure}

It is known that conductivity $\sigma$ behaviour near the
percolation threshold ($ p-p_c \ll 1$) obeys the scaling relations
~\cite{Stauffer}:
 \begin{eqnarray}
  \sigma_ -  \propto (p_c -p)^{-s_c}, & p<p_c   \label{eq01}, \\
  \sigma_ +  \propto (p - p_c)^{t_c},   &  p>p_c   \label{eq02},
 \end{eqnarray}
where $t_c, s_c$ are the conductivity exponents.

A detailed study of the finite size effects is presented in order to discuss
the universality class of the phase transition which the system undergoes. The
main aim of the paper is to determine the dependence

For purposes of checking the universality class and calculating the values of $t_c,s_c$,the
relations \eqref{eq01} and \eqref{eq02} can be written in the following form
\begin{eqnarray}
  \sigma_ - \propto L^{s_c/\nu}, & p<p_c  \label{eq03},\\
  \sigma_ + \propto L^{-t_c/\nu}, & p>p_c  \label{eq04},
\end{eqnarray}
where $L$ is the linear lattice size~\cite{LF84}.

Fig.~\ref{fig:Leb03} shows results of the finite-size analyzes of $\sigma_ -$
and $\sigma_+$. Ratios of the conductivity exponents to the correlation length
exponent $t_c/\nu$ and $s_c/\nu$ are independent of $s$ and close to
$0.973\pm 0.05$, which is a typical value for the percolation in $d=2$~\cite{LF84}.
Hence, our model belongs to the class of random percolation in two dimensions.
This statement agrees with 
the conclusion drawn from analysis of the exponents
$\nu $, $\gamma$, $\beta$.

\begin{figure}[!htbp]
\centering
\includegraphics*[width=0.75\linewidth]{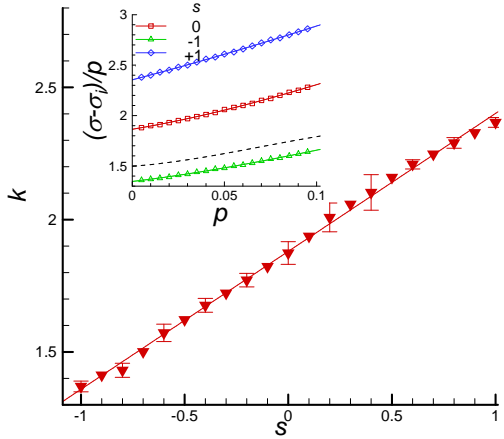}\caption {
Intrinsic conductivity $k$ vs. parameter $s$. In the inset:
$(\sigma/\sigma_i-1)/p$ vs. $p$ for $s =0,\pm 1$ (symbols) and for conventional
random site percolation (dash line). The data correspond to $L \to
\infty$.~\cite{LF84}.}\label{fig:Leb04}
\end{figure}

Near the percolation point, behaviour of the conductivity depends essentially on
$s$. At given $p$, the conductivity can fall down near the percolation threshold
($p\to p_c$) if alignment of dimers ($|s|$ increases). Thus,
alignment induces decrease of the conductivity at given conditions.

However, the situation changes drastically if occupation probability is
rather small $p <0.1$. At given $p$, increase of $s$ leads to decrease of
$\sigma$. The conductivity is maximum when all the dimers are aligned along
the conductivity direction (see left inset in Fig.~\ref{fig:Leb02}). If $p$ is
small enough, the conductivity behaviour can be well fitted by virial
expansion~\cite{Garboczi1995}
\begin{equation}
\sigma/\sigma_i=1+kp+mp^2+\dots,
   \label{eq05}
\end{equation}
where $k$ and $m$ are the adjustable parameters.

The value of intrinsic conductivity $k$ can be extracted from
$(\sigma/\sigma_i-1)/p$ vs. $p$ in the limit of $p \to 0$ (see inset in
Fig.~\ref{fig:Leb04}).

Intrinsic conductivity $k$ as a function of orientation order parameter $s$ is fitted well by linear function $k =1.88+0.52s$ ( with correlation $\rho =0.9976$)
(Fig.~\ref{fig:Leb04}). Notice that intrinsic conductivity calculated for
conventional random site percolation (percolation of monomers) is $1.50 \pm
0.01$ (dash line in inset in Fig.~\ref{fig:Leb04})).

\section{Conclusion}\label{sec:conclusion}

Thus, new percolation problem, i.e. percolation of aligned dimers on a
square lattice, was proposed and studied.
The percolation threshold for deposition of dimers
was lower than for deposition of monomer, $p_c=0.5927\dots$,
nevertheless, the problem belongs to the same universality class.

The lowest percolation threshold $p_c=0.562$ corresponds to isotropic
orientation of the dimers ($s = 0$). In the case of dimers aligned strictly
along one direction ($s=+1,-1$), the percolation point is  $p_c=0.586$. Near
the point of percolation transition ($p\to  p_c$), the conductivity essentially
decreases if the absolute value of orientation order parameter $|s|$ increases.
Intrinsic conductivity $k$ increases linearly with $s$ ($k =1.88 + 0.52s$) and
differs from the known value for a lattice filled with monomers $k=1.50 \pm
0.01$.

The proposed model can be applied to the phase transitions of
anisotropic objects on a lattice when their concentration and
orientation are varied. In particular, the model is useful for
description of a phase transition from insulator to
(semi)conductor upon aligned deposition of the prolate objects on
a substrate. The natural extension of the model is substitution of
dimers by $k$-mers and inclusion of the bonds between the dimers
into consideration.

\section{Acknowledgment}

This work has been supported by Russian Foundation for Basic Research (grant
no.~09-02-90440) and Ministry of Education and Science of Ukraine (Project
no~F28.2/058).


\begin{thebibliography}{99}

\bibitem{Grimmet}
G. Grimmet, \emph{Percolation} (Berlin: Springer-Verlag, 1999)

\bibitem{Kesten}
H. Kesten, \emph{Percolation theory for mathematicians} (Boston: Birkhauser,
1982)

\bibitem{Feder}
J. Feder, \emph{Fractals} (New York: Plenum Press, 1988)

\bibitem{Ziman}
J.M.~Ziman, \emph{Models of disorder. The Theoretical Physics of Homogeneously
Disordered Systems} (Cambridge University Press, 1979)

\bibitem{Shklovskii}
B.I.~Shklovskii, A.L.~Efros, \emph{Electronic properties of doped
semiconductors} (Springer, Heidelberg, 1984)

\bibitem{Stauffer}
D.~Stauffer, A.~Aharony, \emph{Introduction to Percolation Theory}
(Taylor \& Francis, 1992)

\bibitem{Sahimi}
I. Sahimi, \emph{Application of Percolation Theory} (London: Taylor
{\&}Francis, 1994)

\bibitem{Du}
F. Du, J.E.~Fischer, K.I.~Winey, Phys. Rew. B \textbf{72}, 121404 (2005)

\bibitem{Carroll}
 D.~L. Carroll, R.~Czerw, S.~Webster, Synthetic Metals \textbf{155}, 694 (2005)

\bibitem{Sreekumar}
T.V.~Sreekumar, T.~Liu, S.~Kumar, L.M.~Ericson, R.H.~Hauge, R.E.~Smalley, Chem.
Mater \textbf{15}, 175 (2003)

\bibitem{Zhou}
Y.~Zhou, A.~Gaur, S-H.~Hur, C.~Kocabas, M.A.~Meitl, M.~Shim, J.A.~Rogers, Nano
Letters \textbf{4}, 1643 (2004)

\bibitem{dovzhenko}
A.Yu.~Dovzhenko, V.A.~Bunin, Technical Physics \textbf{48}, 123 (2003)

\bibitem{Kondrat}
G. Kondrat, A. P\c{e}kalski, Phys. Rev. E \textbf{63}, 051108 (2001)

\bibitem{Lebovka}
N.V.~Vygornitskii, L.N.~Lisetskii, N.I.~Lebovka, Colloid Journal \textbf{69},
597 (2007)

\bibitem{Liu}
X. Liu, J.L.~Spencer, A.B.~Kaiser, W.M.~Arnold, Current Applied Phys.
\textbf{4}, 125 (2004)

\bibitem{Park}
C. Park, J. Wilkinson, S. Banda, Z. Ounaies, K.E.~Wise, G. Sauti, P.~T.
Lillehei, J.S.~Harrison, Journal of Polymer Science Part B: Polymer Physics
\textbf{44}, 1751 (2006).

\bibitem{Evans}
J.W.~Evans, Rev. Mod. Phys \textbf{65}, 1281 (2003)

\bibitem{Adamczyk}
Z. ~Adamczyk,  \emph{Particles at Interfaces: Interactions,
Deposition, Structure} (Academic Press, 2006)

\bibitem{Holloway}
H. ~Holloway, Phys. Rev. B \textbf{37}, 874 (1988)

\bibitem{Evans1989}
J.W. ~Evans, D.E. ~Sanders, Phys. Rev. B \textbf{39}, 1587 (1989)

\bibitem{Leroyer}
Y. ~Leroyer, E. ~Pommiers, Phys. Rev. B \textbf{50}, 2795 (1994)

\bibitem{Vandewalle}
N. Vandewalle, S. Galam, M. Kramer, Eur. Phys. J. B \textbf{14},
407 (2000)

\bibitem{Cortes}
J. ~Cortes, E. ~ Valencia, J. Colloid \& Interface Sci.
\textbf{252}, 256 (2002)

\bibitem{Cornette1}
V. ~Cornette, A. J. ~Ramirez-Pastor, F. ~Nieto, Physica A
\textbf{327}, 71 (2003)

\bibitem{Cornette2}
V. ~Cornette, A. J. ~Ramirez-Pastor, F. ~Nieto, Eur. Phys. J. B
\textbf{36}, 391 (2003)

\bibitem{Cornette3}
V. ~Cornette, A. J. ~Ramirez-Pastor, F. ~Nieto, Physics Letters
A\textbf{353},452 (2006)

\bibitem{Quintana}
M. ~Quintana, I. ~Kornhauser, R. ~Lopez, A.J. ~Ramirez-Pastor, G.
~Zgrablich, Physica A \textbf{361} 195 (2006).

\bibitem{Dolz}
M. ~Dolz, F. ~Nieto, A.J. ~Ramirez-Pastor, Physica A \textbf{374} 239 (2007)

\bibitem{Pruessner}
G.~Pruessner, M.~Moloney, J. Phys. A \textbf{36}, 11213 (2003)

\bibitem{Matsumoto}
M. Matsumoto, ACM Trans. on Modeling and Computer Simulations \textbf{8}, 3
(1998)

\bibitem{Matoz-Fernandez}
D.A.~Matoz-Fernandez, D.H.~Linares, A.J.~Ramirez-Pastor, EPL \textbf{82}, 50007
(2008)

\bibitem{Hoshen}
J.~Hoshen, R.~Kopelman, Phys. Rev. B \textbf{14}, 3438 (1976)

\bibitem{FL88}
D.J.~Frank , C.J.~Lobb, Phys. Rev. B \textbf{37}, 302 (1988)

\bibitem{Yuge78}
Y. Yuge , K. Onizuka, J. Phys. C: Solid State Phys. \textbf{11}, 4095 (1978)


\bibitem{Bunde}
A. Bunde, S. Havlin, \emph{Fractals and Disordered Systems, eds. A. Bunde and
S. Havlin} (Springer, 1996)

\bibitem{LF84}
C.J.~Lobb , D.J.~Frank, Phys. Rev. B \textbf{30}, 4090 (1984)

\bibitem{Garboczi1995}
E.J.~Garboczi , K.A.~Snyder, J.F.~Douglas, Phys. Rev. E \textbf{52}, 819 (1995)

\end{thebibliography}
\end{document}